# Machine learning for adjoint vector in aerodynamic shape optimization

Mengfei Xu[1], Shufang Song[1], Xuxiang Sun[1], Wengang Chen[1], Weiwei Zhang[1]

**Abstract**     Adjoint method is widely used in aerodynamic design because only once solution of flow field is required for adjoint method to obtain the gradients of all design variables. However, the calculation cost of adjoint vector is approximately equal to that of flow computation. In order to accelerate the solution of adjoint vector and improve the adjoint-based optimization efficiency, machine learning for adjoint vector modeling is presented. Deep neural network (DNN) is employed to construct the mapping between the adjoint vector and the local flow variables. DNN can efficiently predict adjoint vector and its generalization is examined by a transonic drag reduction about NACA0012 airfoil. The results indicate that with negligible calculation cost of the adjoint vector, the proposed DNN-based adjoint method can achieve the same optimization results as the traditional adjoint method.

✉ Shufang Song

shufangsong@nwpu.edu.cn

[1] Northwestern Polytechnical University, Xi 'an, 710072 China

# 1. Introduction

Gradient-based optimization algorithm has become the most widely used engineering optimization algorithm because of its high efficiency. The calculation cost of adjoint method is irrelevant to the number of design variables[1], and is merely proportional to the number of objective functions. In many design optimization problems, the number of objective functions is much lower than the number of design variable. Therefore, the adjoint-based optimization method is efficient and has been fully developed in the field of aerodynamic design. Pironneau[2] adopted the adjoint method in solving the aerodynamic shape optimization problem under Stokes flow. Jamson[3, 4] first proposed applying the continuous adjoint method to the aerodynamic shape optimization problem of transonic velocity range. Later, Nielsen and Anderson[5] proposed the discrete adjoint method. Compared with the continuous adjoint equations that derived from the continuous governing equations, the discrete adjoint equations are derived from the discrete governing equations, and can be solved in the same approach as the governing equations. Since the continuous adjoint equations need to be discretized for numerically solving, the solution of discrete adjoint equations is of higher accuracy and thus adopted by most researchers. Recently, Lyu et al.[6] employed the discrete adjoint method based on the Spalart-Allmaras turbulence model equation in the single-point and multipoint optimization of a NASA Common Research Model wing. Chen et al.[7, 8] studied the optimization problems of flutter suppression on the aircraft control surface and vortex shedding suppression in the flow past a blunt body, and improved the stability of fluid-structure interaction system by discrete adjoint method. He et al.[9] achieved aerothermal optimization for design variables of a U-shaped internal cooling passages based on discrete adjoint method. Kenway et al.[10] summarized the historical research developments and applications of adjoint method in computational fluid dynamics (CFD), studied and compared the various approaches to implement discrete adjoint method.

The current research is mainly focused on the applications of adjoint method. However, although the calculation cost of adjoint vector is irrelevant to the number of design variables, it is still approximately equal to once flow-field computation. Therefore, the calculation cost cannot be ignored in the optimization process. To reduce the calculation cost, Li et al.[11] constructed a reduced-order model (ROM) of the adjoint vector via proper orthogonal decomposition, and enabled the aerodynamic design optimization of RAE2822 and ONERA M6 wings by the ROM. The calculation time of the adjoint vector by the ROM is only about 30% of the traditional adjoint method (TAM). Chen et al.[12] considered the iterative solution of adjoint vector as a dynamic process, and reduced the iteration steps of adjoint vector by 60%. The above methods can only save part of the solution time of the adjoint vector. If the adjoint vector can be predicted only by few samples, the efficiency of aerodynamic shape optimization will be remarkably improved, which is of great significance in the engineering field.

In recent years, artificial intelligence has achieved outstanding performance beyond human in the fields of natural language processing[13] and image recognition[14, 15]. Data-driven methods are widely used in the construction of equations and models. Machine learning methods, such as artificial neural network (ANN), have been widely applied and studied in the field of fluid mechanics[16]. The study of neural network appeared as early as 1989, Hornik et al.[17] stated that any function can be approximated by a sufficiently large and deep neural network. Recently, there is also theory[18] indicated that sparsely connected DNN can be information theory-optimal nonlinear approximator for various functions and systems. There are many applications of ANN in the field of fluid mechanics, such as in machine learning methods for turbulence modeling, Zhu et al.[19] used a single-layer radial basis neural network (RBFNN) to constructed a turbulent algebraic model, and realized the coupling calculation with the CFD solver. Ling et al.[20] constructed a deep neural network with tensor bases as input features to predict the Reynolds stress of typical flow at low Reynolds numbers (Re). Beck et al.[21] employed the residual convolution neural network to close the stress term in large eddy simulations and realized the coupling calculation with the CFD solver. In addition, related applications in fluid mechanics include the identification of ordinary differential equations[22] and partial differential equations[23, 24], etc. In the field of aerodynamic shape optimization, some researchers constructed surrogate model based on ANN. Su et al.[25] integrated RBFNN ensemble and mesh adaptive direct search algorithm, and proved the less calculation cost compared with traditional genetic algorithm (GA) via an example of airfoil optimization. Pehlivanoglu et al.[26] used

RBFNN as a local surrogate model and combined it with GA in the multi-element airfoil optimization. Kharal et al.[27] employed the fully connected neural network to construct the mapping between the aerodynamic coefficients at different angles of attack and airfoil parameters at fixed Mach number and Re, and realized the inverse design of the airfoil. Sekar et al.[28] used convolutional neural network to construct a mapping between pressure distribution and airfoil, and realized the inverse design of the airfoil. However, this method requires up to 1200 airfoils and corresponding pressure distributions as training samples even at the fixed angle of attack and Re, which is difficult to meet in actual industry. Researchers also carried out DNN in aerodynamic shape optimization, such as Yan et al.[29, 30] built a DNN with the design variables of the rocket as the input and the corresponding changes as the output. However, there are strict limitations on design variables and shape parameterization in this modeling strategy.

The above machine learning methods for aerodynamic shape optimization always require massive amounts of samples and lack of generalization of incoming state. In addition, there is a lack of effective surrogate model to predict adjoint vector. Therefore, DNN is presented to construct the mapping between flow field information and adjoint vector. The optimization is accelerated by replacing the solution of adjoint equations with the DNN-based adjoint method (DAM).

The structure of paper is as follows. In Sect. 2, the basic theory of CFD numerical methods and adjoint method are briefly introduced. In Sect. 3, the framework of DNN and the modeling workflow of DAM are introduced firstly, and in subsection 3.3 the impact of different input features on the accuracy of modeling is studied. In Sect. 4, the drag reduction results of DAM are compared with that of TAM in the test state. Finally, Sect. 5 summarizes the paper.

## 2. Numerical methods

### 2.1 Governing equations

As a preliminary attempt of machine learning for adjoint vector modelling, Euler's equation is used as the governing equation to explore the feasibility of proposed method.

In the two-dimension cartesian coordinate system, when the volume force and heat source are ignored, the integral form of steady Euler's equation is

$$\oint_{\partial \Omega} F(U) \cdot \vec{n} d\Gamma = 0 \tag{1}$$

where $U$ is the vector of conservative variables, $\Omega$ is the control volume, $F(U)$ is the vector of viscous fluxes. $\Gamma$ represents the boundary of the control volume $\Omega$, $\vec{n}$ donates the unit outward normal vector to the boundary.

$$U = \begin{Bmatrix} \rho \\ \rho v \\ \rho E \end{Bmatrix}, F(U) = \begin{Bmatrix} \rho v \\ \rho v v^T + p \\ (p + \rho E) v \end{Bmatrix} \tag{2}$$

where $\rho$ is the density, $v$ is the velocity vector, $E$ is the total energy, and $p$ is the pressure. Here we need to introduce the ideal gas equation to close the Eq.(1):

$$p = (\gamma - 1)\rho \left( E - \frac{|v|^2}{2} \right) \tag{3}$$

Euler's equation is discretized according to the finite volume method, then Eq.(1) on the control volume $V_i$ within boundaries can be written as,

$$\sum_{j=1\ldots N_i} \Phi_{ij} = 0 \tag{4}$$

where $N_i$ represents the number of boundaries around the control volume $V_i$. Let $r_i = \sum_{j=1\ldots N_i} \Phi_{ij}$, we can get

$$R = 0 \tag{5}$$

where $R = [r_1, r_2, \ldots, r_N]^T$ refers to spatial discrete terms, and $N$ is the total number of control volumes.

## 2.2 Adjoint method

After the calculation of flow field is completed, we can obtain the objective function $J$ according to the conservative variables $U$ and coordinates of grid points $x_{grid}$,

$$J = J(U, x_{grid}) \tag{6}$$

When the design variables $D$ change, they will affect the value of the objective function $J$ by affecting the conservative variables $U$ and the coordinates of the grid points $x_{grid}$. Therefore, the derivative of the objective function with respect to design variables can be obtained according to chain derivation rule as follows

$$\frac{dJ}{dD_j} = \frac{\partial J}{\partial U}\frac{\partial U}{\partial D_j} + \frac{\partial J}{\partial x_{grid}}\frac{\partial x_{grid}}{\partial D_j}, (j=1\ldots N_D) \tag{7}$$

Equation (7) can be shortened to,

$$\frac{dJ}{dD_j} = \frac{\partial J}{\partial D_j} + \frac{\partial J}{\partial U}\frac{\partial U}{\partial D_j}, (j=1\ldots N_D) \tag{8}$$

where $N_D$ is the number of design variables. $\partial J/\partial D_j$ and $\partial J/\partial U$ on the right-hand side are easy to obtain. However, there is difficulty in calculating $\partial U/\partial D_j$. Therefore, we introduce the adjoint vector $\Lambda$, which can be solved by,

$$\frac{\partial R^T}{\partial U}\Lambda = \frac{\partial J^T}{\partial U} \tag{9}$$

Transposing the above formula and multiplied by $\partial U/\partial D_j$

$$\Lambda^T \frac{\partial R}{\partial U}\frac{\partial U}{\partial D_j} = \frac{\partial J}{\partial U}\frac{\partial U}{\partial D_j} \tag{10}$$

Considering and linearizing the governing equation $R = 0$, then

$$\frac{\partial R}{\partial U}\frac{\partial U}{\partial D_j} = -\frac{\partial R}{\partial D_j} \tag{11}$$

Substituting Eq.(11) into Eq.(10)

$$\frac{\partial J}{\partial U}\frac{\partial U}{\partial D_j} = -\Lambda^T \frac{\partial R}{\partial D_j} \tag{12}$$

Substituting Eq.(12) into Eq.(8), we can get

$$\frac{dJ}{dD_j} = \frac{\partial J}{\partial D_j} - \Lambda^T \frac{\partial R}{\partial D_j}, \ (j=1\ldots N_D) \tag{13}$$

Therefore, once the adjoint vector $\Lambda$ is obtained via Eq.(9), the gradients of all design variables can be obtained at negligible cost via Eq.(13). The original problem of solving $N_D$ linear equations is transformed into solving only one linear equation. However, the solution of $\Lambda$ is still time consuming. Considering the definition of $\Lambda$ in Eq.(9),

$\partial \boldsymbol{R}^T / \partial \boldsymbol{U}$ is a large sparse matrix of size $N_e \times N_e$, and $\partial J^T / \partial \boldsymbol{U}$ is a vector of size $N_e \times 1$. $N_e$ is the total number of grid points, which can be more than $10^4$ in two-dimension case. TAM requires multi-step iteration to solve such a large sparse linear system. However, through DNN, the adjoint vector will be predicted quickly based on the flow field information. Then the gradients can be calculated via Eq.(13), so as to efficiently optimize aerodynamic shape. The accuracy verification of flow field solution and adjoint method program can be referred to Chen et al.[12].

## 3. Machine learning for adjoint vector

### 3.1 Deep neural network

As one of the most well-known method in machine learning, DNN possesses a strong capability to approximate nonlinear system, which benefits from its modular structure as shown in Fig. 1. DNN is composed of multiple neural layers, each with many neurons, and data is transmitted between layers through connection lines with weights. The number of neurons in the input layer depends on the number of features of the flow field on the grid points, while the number of neurons in the output layer corresponds to the number of adjoint vector's components.

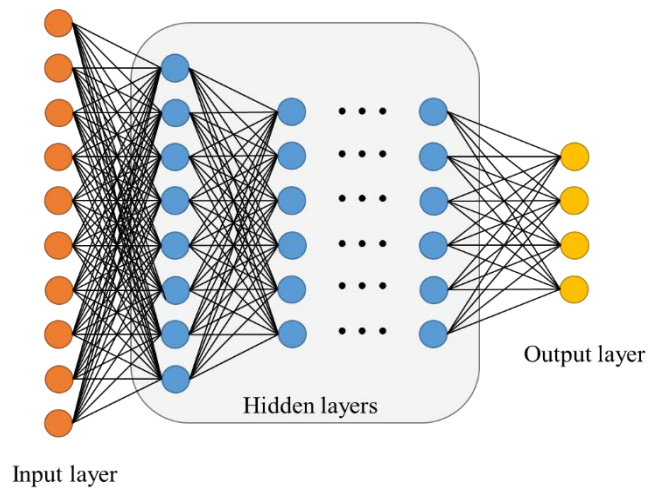

Fig. 1 Framework of deep neural networks

The input layer receives the data of flow-field features, and the sum of weighted data and bias is transferred to the next layer. Activation function is performed on each hidden layer. Take the first hidden layer as an example, the calculation from the inputs to the first hidden layer neuron can be expressed as

$$Y_j(X) = \sigma\left(\sum_{i=1}^{N} \omega_{j,i} X_i + b_j\right) \tag{14}$$

where $X$ indicates the input data, $\omega_{j,i}$ represents the weight connecting the input layer neuron $i = (1, 2, \ldots, N)$ and the first hidden layer neuron $j = (1, 2, \ldots, S)$. $N$ and $S$ are the number of the neurons in the input layer and hidden layer respectively. $b$ refers to the bias vector. $\sigma$ indicates the activation function in the hidden layer. Common choices of activation function are nonlinear functions such as ReLU, Sigmoid and Tanh.

Finally, the data is transferred to the output layer and produces a prediction of adjoint vector. Once the loss function is determined, the partial derivatives of loss function with respect to the learnable parameters $\omega$ and $b$ can be obtained according to chain derivation rule. In order to achieve the minimum value of the loss function, the parameters will be updated via the error backpropagation algorithm and optimization algorithm (such as Adam[31]) based on the partial derivatives. In this paper, the pytorch library[32] is used to build and train the DNN, and the adopted loss function is mean squared error loss (MSELoss). Through the nonlinear transformation in multiple hidden layers, the data of flow-field features is mapped to a space on which the distribution is consistent with the adjoint vector.

In general, the deeper the hidden layer is, the stronger the power of DNN is. However, as the hidden layers deepen, the training becomes complicated due to the phenomenon of *gradient disappearance* and *gradient explosion*. This phenomenon can be avoid by using Batch Normalization method[33], which normalizes layer inputs, accelerates training effectively and improves the accuracy. Also, in order to facilitate training, the inputs and outputs of DNN are linearly scaled in [0,1] interval.

The overall hyperparameters of the constructed DNN is briefly described as follows. The DNN is composed of 5 hidden layers, of which the number of layer neurons is 100, 100, 80, 50 and 30 respectively. Batch Normalization method is adopted for all hidden layers. In the output layer, there are 4 neurons corresponding to the four components of the adjoint vector. The training is divided into two stages. In the first stage, the batch size, the initial learning rate and the number of training epochs is set to $2^{15}$, 0.03 and 1000, and the second stage to $2^9$, 0.003 and 600 respectively. During the training process, when no decrease of minimum value of validation loss is seen for 40 training epochs, the learning rate is halved.

## 3.2 Modeling workflow

The specific modeling workflow is shown as Fig. 2. The class-shape-transform (CST) methodology[34] is used to parameterize the geometric shape of airfoil. Samples are calculated by CFD solver at different boundary conditions of CST parameters, Mach number ($Ma$) and angle of attack ($\alpha$). The adjoint vector $\Lambda$ can be obtained by adjoint solver base on the flow-field solution $(\rho, u, v, p)$. A data set is composed of data of flow-field features $(\rho, u, v, p, ...)$ as inputs and $\Lambda$ as outputs for training and validation. The proposed DNN is trained on the data set to locally map from $(\rho, u, v, p, ...)$ to $\Lambda$. Therefore, the iterative solution of adjoint vector by TAM can be replaced by predicting $\hat{\Lambda}$ via DNN based on the data of $(\rho, u, v, p, ...)$. Finally, the estimated gradients can be obtained.

There are two main advantages of this approach. First of all, the training benefits from the massive amounts of data in CFD. Naturally, each flow-field solution is composed of tens of thousands of values on mesh cell. Secondly, the local mapping possesses a certain generalization and is irrelevant to the computational mesh.

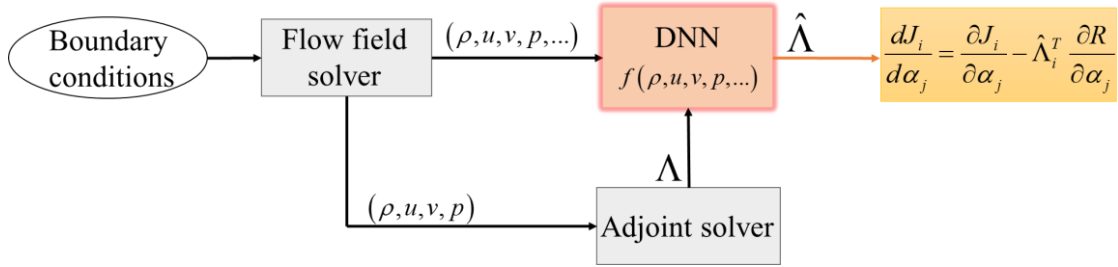

Fig. 2. Flow chart of DNN-Based adjoint method

### 3.3 Input features design

According to the definition of $\Lambda$ in Eq.(9), when the objective function $J$ is determined, $\partial \boldsymbol{R}^T/\partial \boldsymbol{U}$ and $\partial J^T/\partial \boldsymbol{U}$ can be written as a function of the whole-field $(\rho, u, v, p)$. In consequence, $\Lambda$ are also uniquely determined by the whole-field. However, the constructed DNN maps from the local $(\rho, u, v, p, ...)$ to the local $\Lambda$, which is incomplete in physics. In order to address this problem, the spatial gradients of $(\rho, u, v, p)$ are added to the inputs of DNN, as shown in Table 1. The incompleteness of mapping can be compensated to a certain extent by this strategy. In addition, the accuracy of DNN is also improved.

Table 1 Design of flow-field features

| Basic flow features | $(\rho, u, v, p)$ |
|---|---|
| Features including spatial gradients | $\left(\rho, u, v, p, \dfrac{\partial \rho}{\partial x}, \dfrac{\partial \rho}{\partial y}, \dfrac{\partial u}{\partial x}, \dfrac{\partial u}{\partial y}, \dfrac{\partial v}{\partial x}, \dfrac{\partial v}{\partial y}, \dfrac{\partial p}{\partial x}, \dfrac{\partial p}{\partial y}\right)$ |

In order to verify the modeling capability of features, training data of the NACA0012 is obtained at $Ma = 0.8$, $\alpha = 3°$ with drag coefficient $C_d$ as the objective function. DNN is trained with the basic flow features as inputs and features including spatial gradients as inputs respectively. The absolute mean error is 0.3158 and 0.1775 respectively. In Fig. 3, the contour images of the first component of $\hat{\Lambda}$ is compared with the ground truth. By adding spatial gradients to the inputs, the accuracy of the adjoint vector modeling is significantly improved.

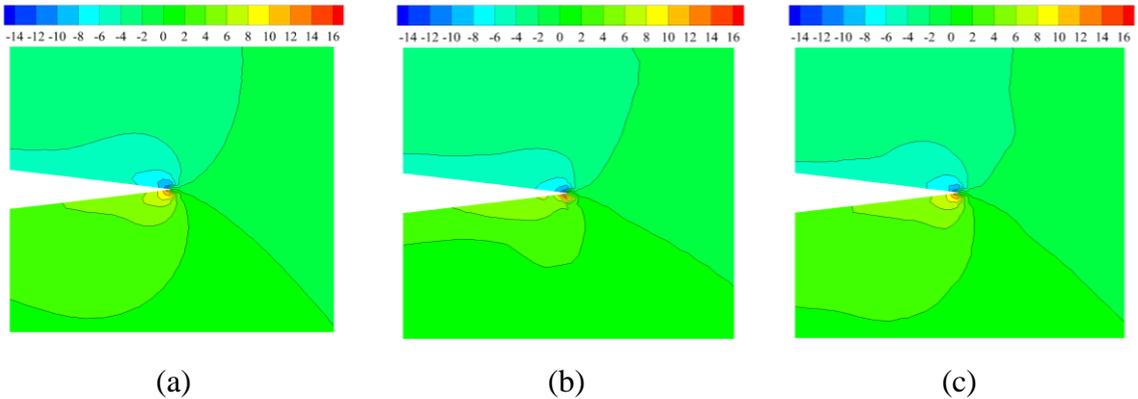

(a)　　　　　　　　　　(b)　　　　　　　　　　(c)

Fig. 3 Comparison of first component of $\Lambda$. **a** Ground truth. **b** By basic flow features. **c** By features including spatial gradients.

Furthermore, DNN of the above architecture is used to learn the mapping between the data of flow-field features including spatial gradients and $\Lambda$ in two cases. As shown in Table 2, data of training is obtained at subsonic incoming state with lift coefficient $C_l$ as the objective function in the first case, and at transonic incoming state with $C_d$ as the objective function in the second case. Then $\hat{\Lambda}$ and the estimated gradients produced by the DNN are compared with the ground truth.

Table 2 Information of two cases

|  | **Airfoil** | *Ma* | $\alpha$ | **Objective Function** |
|---|---|---|---|---|
| **Case1** | NACA0012 | 0.5 | 3° | $C_l$ |
| **Case2** | NACA0012 | 0.75 | 2° | $C_d$ |

The contour image comparisons of $\Lambda$ and CST gradients on case1 and case2 are illustrated in Fig. 4 and Fig. 5, respectively. The first four contour images correspond to the four components of $\Lambda$. From the comparisons, it is incredibly accurate for adjoint vector modeling via DNN in training cases, regardless of subsonic or transonic incoming states and $C_l$ or $C_d$ as objective functions. From Fig. 4(a-d) and Fig. 5(a-d), the contour images of $\hat{\Lambda}$ are almost the same distribution as that of $\Lambda$, which is calculated by TAM. The gradients of CST obtained from $\Lambda$ and $\hat{\Lambda}$ match well as shown in Fig. 4(e) and Fig. 5(e).

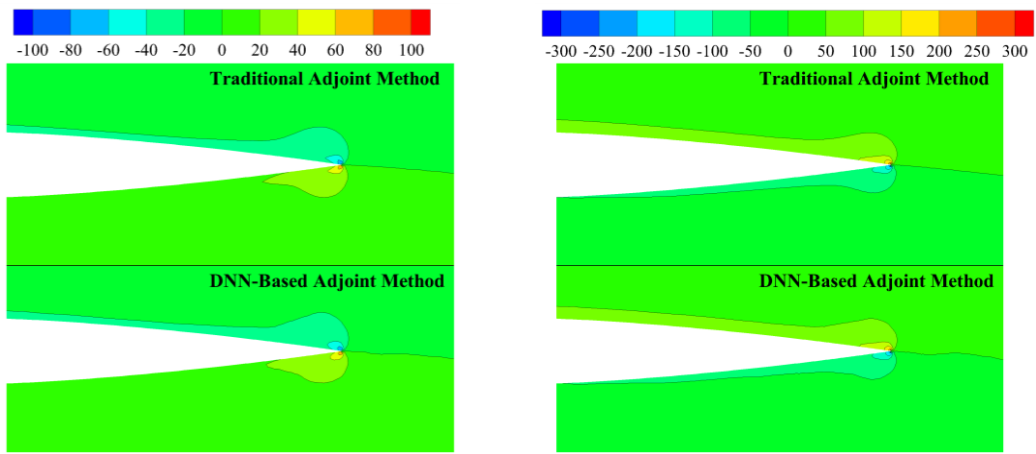

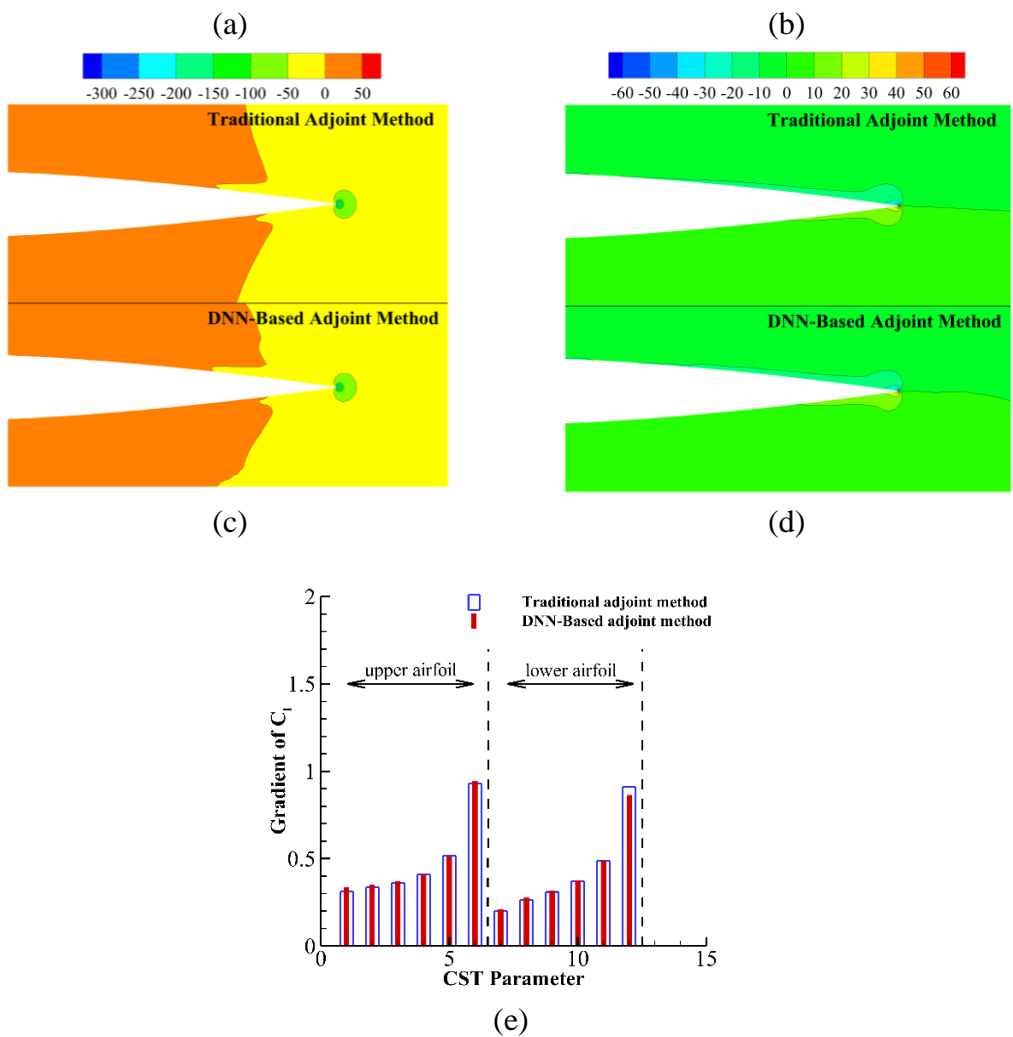

Fig. 4 Comparisons of the traditional adjoint method and DNN-Based adjoint method on case1. **a** First component of $\Lambda$. **b** Second component of $\Lambda$. **c** Third component of $\Lambda$. **d** Fourth component of $\Lambda$. **e** Gradients of 12 CST parameters.

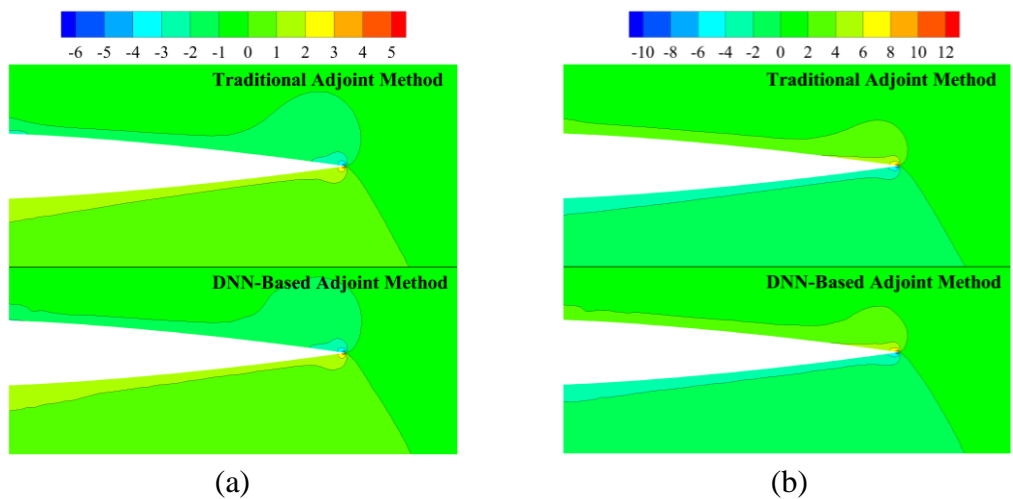

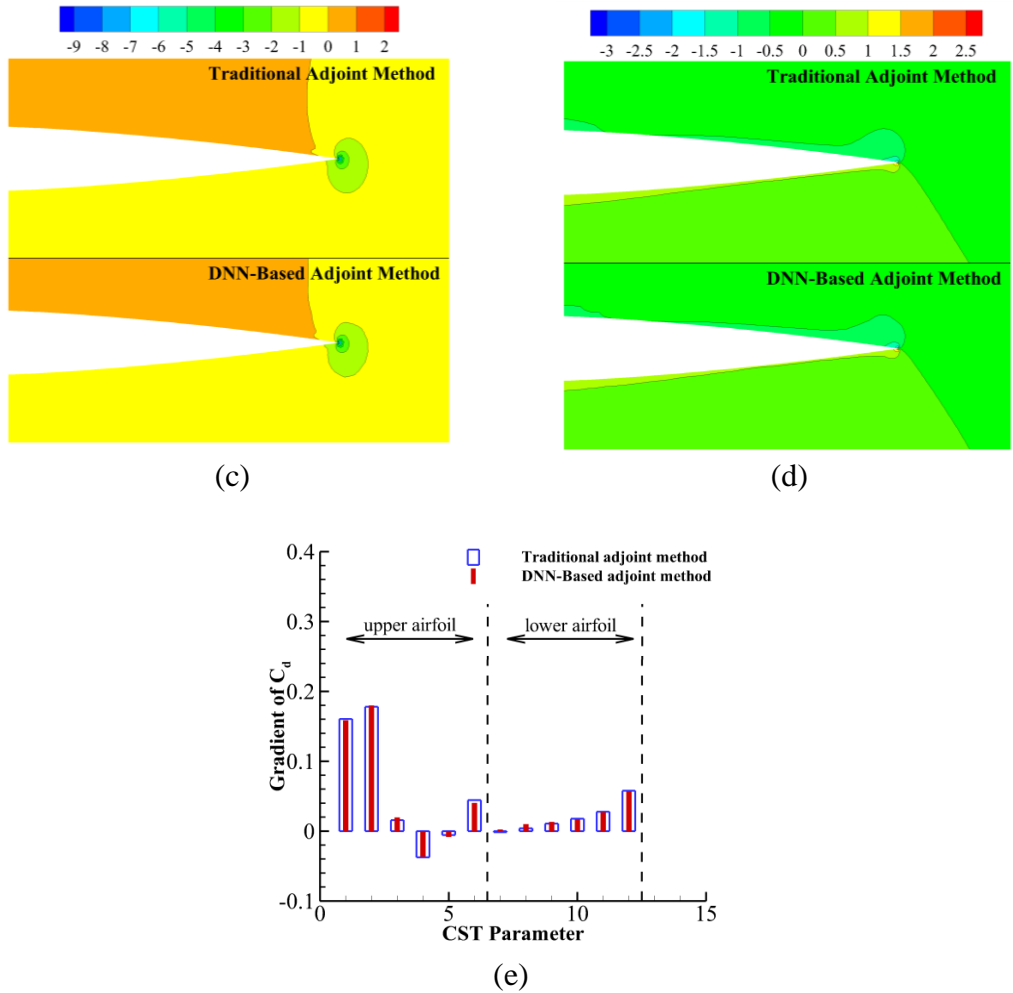

Fig. 5 Comparisons of the traditional adjoint method and DNN-Based adjoint method on case2. **a** First component of $\Lambda$. **b** Second component of $\Lambda$. **c** Third component of $\Lambda$. **d** Fourth component of $\Lambda$. **e** Gradients of 12 CST parameters.

# 4. Drag reduction about NACA0012 airfoil by DNN-based adjoint method

## 4.1 Analysis of required sample number at fixed incoming state

The powerful fitting capability of the DNN has been proved in the previous section. Next, we study on how the number of samples for training and validation affects the generalization of the DNN. The upper surface and lower surface of the airfoil are both parameterized by 6 CST parameters. By Latin-hypercube sampling[35] within the variation ±30% of CST parameters of NACA0012 airfoil, 30 sample airfoils are randomly selected. All sample airfoils are different from NACA0012 airfoil, as shown in Fig. 6.

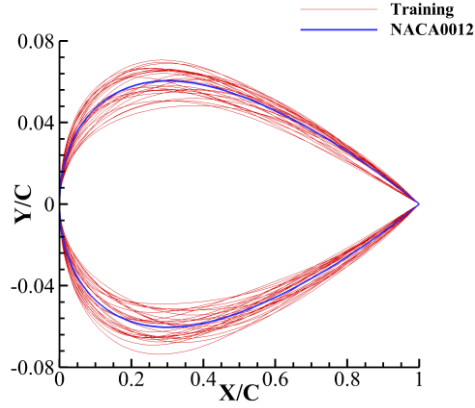

Fig. 6 Distribution of sample airfoils

The data of flow-field features and $\Lambda$ of all sample airfoils is obtained by CFD solver at $Ma = 0.78$, $\alpha = 2.5°$, and then added successively to the data set for training and validation. 20% of the data is randomly selected as the validation set, and the remaining 80% as the training set. During the training process, the validation loss, which is calculated on the validation set, will not be backpropagated to identify the network weights and biases, but only as a reference results to prevent overfitting. Due to the randomness in training-validation-spilt process, batch selection during training and the initialization of model parameters, the training process is repeated 5 times. The minimum test loss is regarded as the performance of DNN. Fig. 7 illustrates the function between the minimum test loss and the number of sample airfoils. This curve can be well fitted by the power function:

$$MSELoss = 3.584 \times 10^{-4} n^{-1.336} + 8.884 \times 10^{-5}$$

where $n$ indicates the number of sample airfoils for training and validation. It can be seen from Fig. 7 that when the number of sample airfoils increases from 1 to 5, the minimum test loss decreases significantly. After that, as the number of sample airfoils increases, the minimum test error decreases more slowly. Therefore, at fixed incoming state, it is a reasonable choice considering accuracy and cost to randomly sample 10 airfoils as training and validation set.

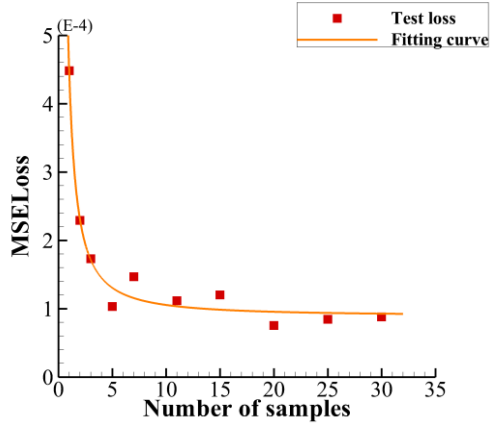

Fig. 7 Minimum MSE Loss as a function of the number of sample airfoils for training and validation

### 4.2 Analysis of prediction accuracy

In the previous section, we explored the relationship between the generalization and the number of sample airfoils at the fixed incoming state. In this section, the generalization of the DNN is studied at the different incoming state. By Latin-hypercube sampling within the range of $Ma = 0.75 \sim 0.78$, $\alpha = 2° \sim 3°$, and ±30% variation CST parameters of NACA0012 airfoil, 30 sample combinations are randomly selected. The data of flow-field features and $\Lambda$ at different combinations is obtained as training and validation set. The training-validation distributions of incoming states and airfoils are shown in Fig. 8. Data of 5 sample combinations is randomly selected as the validation set, and the remaining data of 25 sample combinations as the training set.

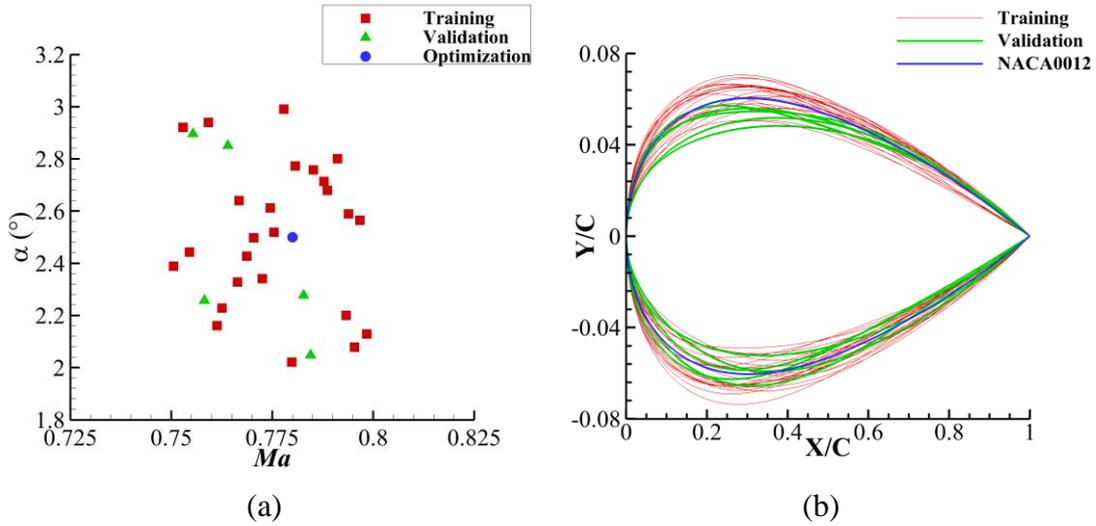

(a)                          (b)

Fig. 8 Distribution of sample combinations. **a** $Ma-\alpha$ distribution. **b** Airfoils distribution.

Fig. 9 shows the convergence of training loss and validation loss during training. The oscillation of validation loss appears around 4e-3 firstly and eventually disappear due to the learning rate reducing. The batch size and learning rate are reset at the $1000^{th}$ epochs of the training process as illustrated above. After a brief oscillation and increase, the training and validation losses eventually converge to a smaller value. By adopting this training strategy, the network weights and biases can be quickly identified. For the explanation, we believe it can speed up training by setting a larger batch size at the beginning. When the networks weights and biases change stably, reducing the batch size appropriately can make the DNN focus more on the learning of the local extremum of $\Lambda$. Therefore, the training and validation losses can be further reduced. During the training process, the DNN achieves the smallest value of validation loss (8.490e-05) at $1247^{th}$ step, while the training loss is 4.237e-05.

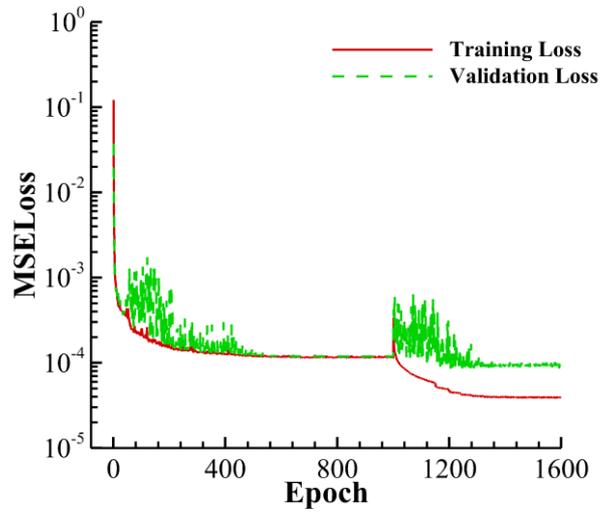

Fig. 9 Training process of DNN

$\hat{\Lambda}$ and gradients of NACA0012 airfoil at $Ma = 0.78$, $\alpha = 2.5°$ are predicted by this trained DNN. Since the combination of NACA0012 airfoil and this incoming state is different from the sample combinations, it is a test case for the DNN. The contour images of $\hat{\Lambda}$ are compared with its ground truth as shown in Fig. 10(a-d). It can be observed that DNN can accurately predict $\Lambda$, even for the extreme values in the tail region. Therefore, the gradients calculated based on the precisely predicted $\hat{\Lambda}$ match well with the TAM results as shown in Fig. 10(e). The error of gradient can be accepted in engineering.

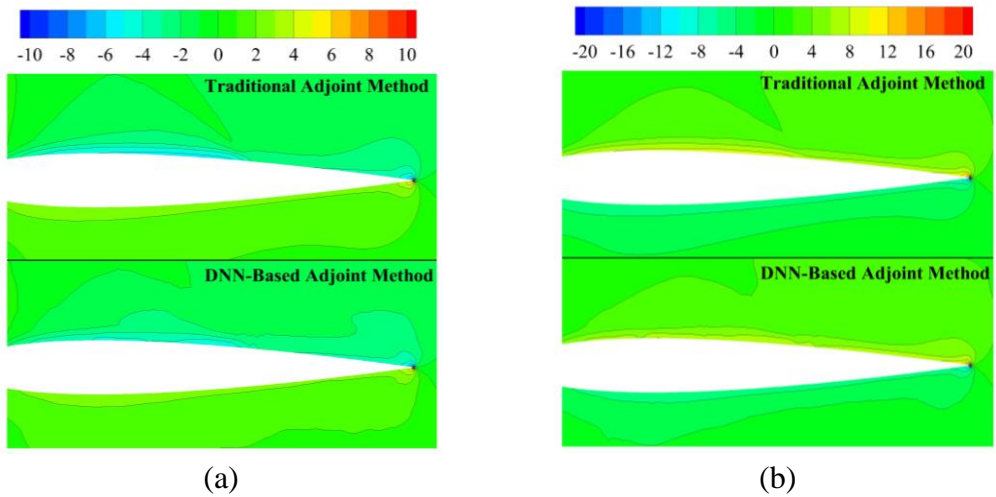

(a)                      (b)

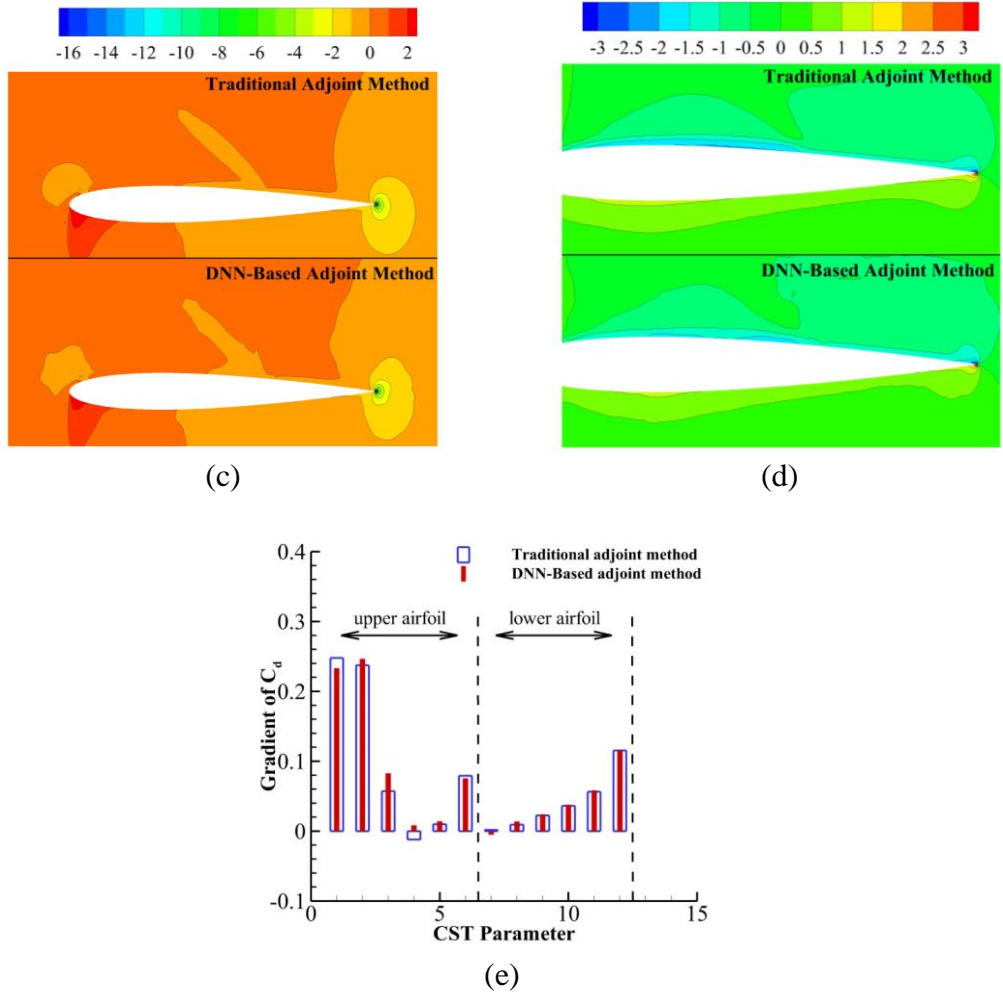

Fig. 10 Comparisons of the traditional adjoint method and DNN-Based adjoint method on NACA0012 at $Ma = 0.78$, $\alpha = 2.5°$. **a** First component of $\Lambda$. **b** Second component of $\Lambda$. **c** Third component of $\Lambda$. **d** Fourth component of $\Lambda$. **e** Gradients of 12 CST parameters.

## 4.3 Aerodynamic shape optimization by DNN-based adjoint method

Through the verification of the gradients on NACA0012 airfoil, the initial direction of optimization is ensured. Next, the DAM is used to guide the drag reduction about NACA0012 airfoil at $Ma = 0.78$, $\alpha = 2.5°$ in which a certain number of constrains must be satisfied. The lift constrain and area constrain are considered as an equality. The change of CST parameter is limited to be within 30%. The initial lift coefficient $C_{l0}$ is 0.5845, and the initial drag coefficient $C_{d0}$ is 0.0379. The

mathematical model of the optimization is expressed as follows:

$$\begin{aligned}&\min C_d(\boldsymbol{D})\\&\text{s.t.}\quad C_l(\boldsymbol{D})\geq C_{l0}\\&\quad\quad Area(\boldsymbol{D})\geq Area_0\\&\quad\quad |\Delta CST|\leq 0.3CST_0\end{aligned} \quad (15)$$

where $\boldsymbol{D}$ represents design variable vector, which is CST parameter vector in this paper. We adopted penalty function method to constrain the optimization problem. The initial penalty factor and the factor expansion is set to 1 and 10, respectively. Details of setting can be found in Ref.[12].

In Fig. 11, the optimization results of DAM with TAM are compared. As shown in Fig. 11(a, b), optimal airfoil and pressure distribution of DAM are almost the same as those of TAM, though there are very slight differences in the lower surface of airfoil. The predicted gradients of the optimal airfoil by DAM basically match the ground truth (Fig. 11(c)). Therefore, it is foreseeable that if the optimization constraints are properly released, DAM can still obtain the same optimized shape as TAM. Fig. 11(d) illustrates the $C_d$ iteration of DAM and TAM, which takes 22 steps and 18 steps to converge respectively. Airfoils with a lower $C_d$ but outside the constraint boundary are first obtained by both methods. Then the airfoil shape is gradually optimized to the inside of the constraint boundary as the penalty factor increases. Finally, the optimal airfoils that satisfies the constrains but with a relatively larger $C_d$ are obtained by both methods.

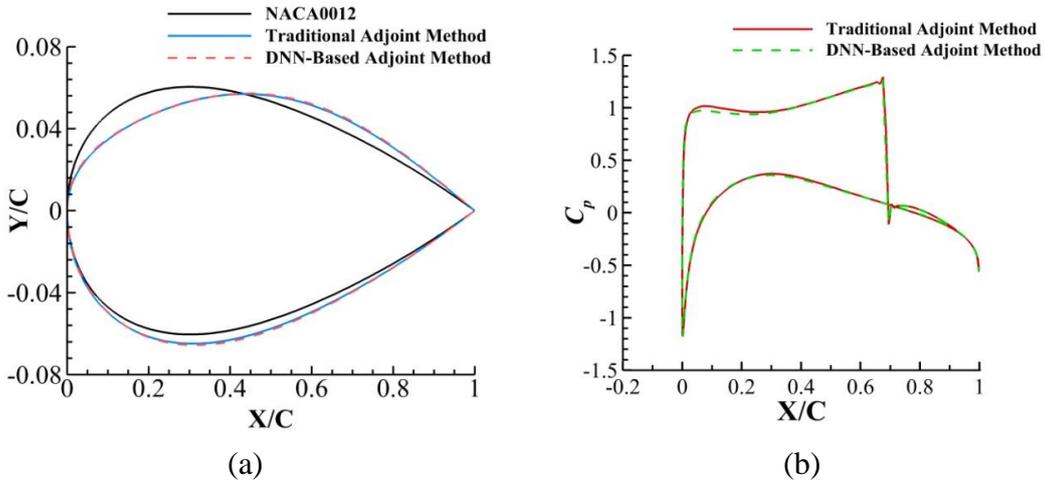

(a)　　　　　　　　　　　　　(b)

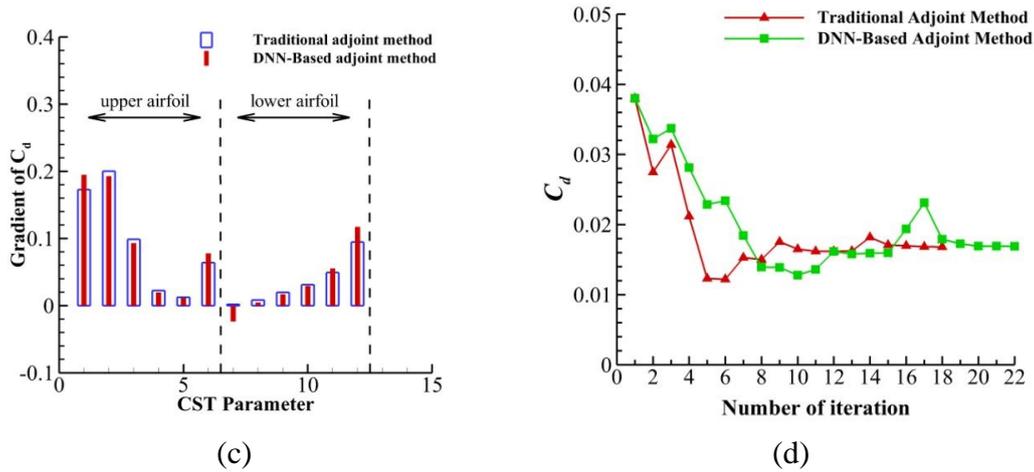

(c)                          (d)

Fig. 11 Optimization results of DNN-based adjoint method and traditional adjoint method at $Ma = 0.78$, $\alpha = 2.5°$. **a** Optimal airfoil. **b** Distribution of $C_p$. **c** Gradients of 12 CST parameters. **d** $C_d$ iteration.

In Table 3, the aerodynamic coefficients of the optimal airfoil by DAM and TAM are compared. The $C_d$ is reduced by 55.6% via TAM and 55.4% via DAM. Therefore, the slight difference in the lower surface of optimal airfoil has little impact on $C_d$, which proves the high accuracy of the DAM.

Table 3. **Results of NACA0012 airfoil and optimal airfoils.**

|  | $C_d$ | $C_l$ | AREA |
| --- | --- | --- | --- |
| NACA0012 | 378.9E-04 | 0.5845 | 0.0821 |
| Optimal airfoil by TAM | **168.1E-04** | 0.5856 | 0.0837 |
| Optimal airfoil by DAM | **168.9E-04** | 0.5969 | 0.0844 |

## 5. Conclusion

DNN is constructed to learn a nonlinear mapping between the local flow field information and the adjoint vector. By adding the spatial gradients of the local flow field to the input features of DNN, the prediction accuracy is significantly improved. The function between the minimum test loss and the number of samples for training and validation is studied and well fitted by a power function. By training on data at different combinations of transonic incoming state and airfoils, DNN possesses the generalization of $Ma$, $\alpha$ and shape. Through the drag reduction about NACA0012 airfoil at the test incoming state, it is verified that the optimal airfoil and the pressure distribution by DAM match well with those by TAM. The results indicate that by learning a small number of samples in a certain range, the DNN can preciously predict the adjoint vector and achieve efficient aerodynamic shape optimization. The following work will be extended to the adjoint vector modelling based on unsteady Navier-Stokes equation.